\documentclass[aps,prl,superscriptaddress,twocolumn,nofootinbib,preprintnumbers]{revtex4}
\pdfoutput=1

\usepackage{color}
\usepackage{graphicx}
\usepackage[space]{grffile}

\usepackage{subfigure}
\usepackage{verbatim}
\usepackage{amsmath}
\usepackage{amssymb}
\usepackage{wasysym}
\usepackage{url}
\usepackage{bbold}
\usepackage{slashed}
\usepackage{epstopdf}

\usepackage{diagbox}
\usepackage{amssymb,amsmath,latexsym,graphics, graphicx,epsfig,multirow,comment,hyperref,appendix,feyn,slashed,xcolor,afterpage, makecell} 
\usepackage{booktabs}
\usepackage{tabularx}
\usepackage{xspace}

\usepackage{multirow}
\usepackage{threeparttable}
\usepackage{paralist}

\usepackage{color}
\definecolor{darkgreen}{rgb}{0,0.5,0}
\definecolor{violet}{rgb}{0.5,0.5,1}

\newcommand{\FeH}{\text{[Fe/H]} }
\newcommand{\alphaFe}{\text{[$\alpha$/Fe]} }

\DeclareRobustCommand{\SFig}[1]{Supplementary Fig.~}
\DeclareRobustCommand{\SFigs}[2]{Supplementary Figs.~}

\newcommand{\be}{\begin{equation}}
\newcommand{\ee}{\end{equation}}
\newcommand{\bea}{\begin{eqnarray}}
\newcommand{\eea}{\end{eqnarray}}
\newcommand{\bi}{\begin{itemize}}
\newcommand{\ei}{\end{itemize}}

\usepackage{xspace}

\providecommand{\eg}{\emph{e.g.}\xspace}

\providecommand{\Gaia}{\emph{Gaia}\xspace}
\providecommand{\Latte}{\emph{Latte}\xspace}
\providecommand{\FIRE}{\textsc{Fire}\xspace}
\providecommand{\RAVE}{RAVE\xspace}
\providecommand{\mi}{\texttt{m12i}\xspace}
\providecommand{\mf}{\texttt{m12f}\xspace}

\providecommand{\Msun}{\,M_{\odot}}

\providecommand{\kpc}{\,\text{kpc}}

\providecommand{\kms}{\,\text{km/s}}

\begin{document}

\title{
{Evidence for a Vast Prograde Stellar Stream in the Solar Vicinity}
}



\author{Lina Necib}
\email{lnecib@caltech.edu}
\affiliation{Walter Burke Institute for Theoretical Physics,
California Institute of Technology, Pasadena, CA 91125, USA \vspace{1.8pt}}

\author{Bryan Ostdiek}
\affiliation{Institute of Theoretical Science, Department of Physics, University of Oregon, Eugene, OR 97403, USA\vspace{1.8pt}}

\author{Mariangela Lisanti}
\affiliation{Department of Physics, Princeton University, Princeton, NJ 08544, USA \vspace{1.8pt}}

\author{Timothy Cohen}
\affiliation{Institute of Theoretical Science, Department of Physics, University of Oregon, Eugene, OR 97403, USA\vspace{1.8pt}}

\author{Marat Freytsis\vspace{2pt}}
\affiliation{Raymond and Beverly Sackler School of Physics and Astronomy, Tel-Aviv University, Tel-Aviv
69978, Israel \vspace{1.8pt}}
\affiliation{School of Natural Sciences, Institute for Advanced Study, Princeton, NJ 08540, USA \vspace{1.8pt}}

\author{Shea Garrison-Kimmel}

\author{Philip F.\ Hopkins}
\affiliation{TAPIR, California Institute of Technology, Pasadena, CA 91125, USA \vspace{1.8pt}}

\author{Andrew Wetzel}
\affiliation{Department of Physics, University of California, Davis, CA 95616, USA \vspace{1.8pt}}

\author{Robyn Sanderson\vspace{5pt}}
\affiliation{Department of Physics and Astronomy, University of Pennsylvania, Philadelphia, PA 19104, USA \vspace{1.8pt}}
\affiliation{Center for Computational Astrophysics, Flatiron Institute, New York, NY 10010, USA \vspace{1.8pt}}

\begin{abstract}

Massive dwarf galaxies that merge with the Milky Way on prograde orbits can be dragged into the disk plane before being completely disrupted.  Such mergers can contribute to an accreted stellar disk and a dark matter disk.  We present Nyx, a vast new stellar stream in the vicinity of the Sun, which provides the first indication that such an event occurred in the Milky Way.  We identify about 90 stars that have coherent radial and prograde motion in this stream using a catalog of accreted stars built by applying deep learning methods to the \Gaia data.  Taken together with chemical abundance and orbital information, these results strongly favor the interpretation that Nyx is the remnant of a disrupted dwarf galaxy.  Further justified by the FIRE hydrodynamic simulations, we demonstrate that prograde streams like Nyx can be found in the disk plane of galaxies and identified using our methods.

\end{abstract}

\maketitle

In the $\Lambda$CDM paradigm, galaxies grow through the accretion of smaller satellites~\cite{1978MNRAS.183..341W}.  Accreted material from these mergers plays an important role in building the Milky Way's stellar halo~\cite{Johnston:1996sb}, as demonstrated by the discovery of numerous tidal streams crisscrossing the sky (see~\cite{2016ASSL..420...87G} for a review).  In certain cases, an infalling satellite can interact dynamically with the stellar disk and be preferentially dragged into the Galactic plane~\cite{1986ApJ...309..472Q, 1996ApJ...460..121W}, leaving behind tidal debris that rotates with the disk.
In this work, we present evidence for a new prograde stellar stream in the vicinity of the Sun, whose interpretation provides the first hint that such a merger occurred in our Galaxy.

A massive satellite falling towards the Milky Way's disk experiences increased tidal friction that slows its orbit and eventually leads to its complete dissolution in the disk plane~\cite{1986ApJ...309..472Q}.  Feeling the pull of the satellite's mass, disk stars lying ahead of the satellite lose angular momentum and fall into more central or ``lower'' orbits.  In contrast, disk stars behind the satellite gain angular momentum and move into ``higher'' orbits.  The net result is an overdensity of disk stars trailing behind the satellite that result in  frictional drag.  Such ``disk dragging'' is most pronounced for massive satellites ($\sim$~$1:10$ mergers) 
moving on prograde orbits.  Observations of the Milky Way's disk stars do allow for such an encounter~\cite{2008MNRAS.389.1041R, 2009ApJ...691.1168H, 2009MNRAS.397...44R, 2009ApJ...703.2275P}.  The fact that accreted stellar and dark disks form from such mergers was first demonstrated using $N$-body simulations~\cite{1986ApJ...309..472Q, 1996ApJ...460..121W, 2008MNRAS.389.1041R, 2009ApJ...703.2275P}, and has since been reproduced in cosmological simulations with baryons~\cite{2009MNRAS.397...44R, 2010JCAP...02..012L, 2014ApJ...784..161P, 2017MNRAS.472.3722G}.  

\begin{figure*}
\centering
\includegraphics[width=0.90\textwidth]{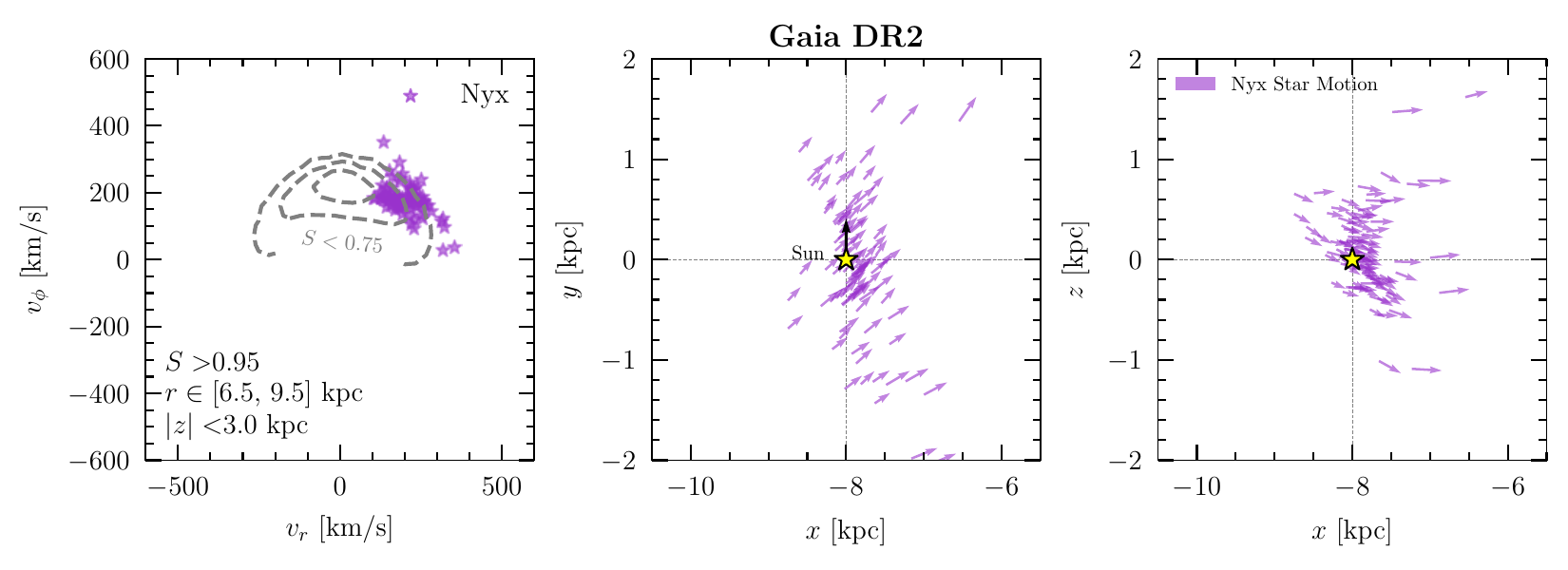}
\caption{\textbf{Dynamics of the Nyx stars.}  (Left) Velocities of the 94 Nyx stars in Galactocentric spherical coordinates $v_r-v_\phi$ in the ROI defined as $r \in [6.5, 9.5]\kpc$ and $|z| < 3\kpc$.
Here, $v_\phi$ rotates with the disk of the Milky Way, and positive $v_r$ points towards the Galactic Center.
We also show 1, 2, and 3$\sigma$ contours in dashed gray for all stars with network scores $S<0.75$.  These are a proxy for \emph{in situ} stars, though they also include accreted stars that the network finds more difficult to classify.  (Middle/Right) Velocity of Nyx stars in the Galactic $x-y$ and $x-z$ plane, where $z$ is orthogonal to the disk. The Galactic Center is located at $(x,y,z) = (0,0,0)\kpc$ while the Sun (indicated by the yellow star moving in the direction of the black arrow) is located at $(x,y,z) = (-8, 0, 0)\kpc$.  The Nyx stream is prograde and passes near the Solar position.  Its kinematics are distinct from the stellar disk, with a rotation speed that lags by $\sim $ 90$\kms$ and a significant radial velocity.  Here, we only show Nyx stars with the highest network scores ($S>0.95$).  As argued in the text, more stars are associated with the stream, but have lower network scores and so are not included here.  \SFig~2 shows the corresponding plot for all stars with $S>0.85$, where we find another potentially related stream, Nyx-2. \SFig~3 shows the corresponding plot, but colored by the score  of each star. 
 }
\label{fig:toomre_095}
\end{figure*}

The primary difficulty in identifying accreted stellar debris near the Galactic midplane is distinguishing it from \emph{in situ} stars born in the disk.  This challenge is amplified when the tidal debris corotates with the disk.  Indeed, standard selection methods for accreted stars conservatively ignore all stars with speeds $|v-v_\text{lsr}| \lesssim 220\kms$, where $v_\text{lsr}$ is the velocity of the local standard of rest. Chemo-dynamical template methods were used to search for an accreted stellar disk, but the analysis was not optimized for prograde structures with similar metallicities to the thick disk ~\cite{Ruchti:2015bja}. 
This motivates the use of techniques that can separate accreted stars from the \emph{in situ} stellar background without significantly compromising statistics. 

To address these issues, we used deep neural networks in Ostdiek et al. 2019 \cite{ml_paper} to build a catalog of accreted stars from the second data release (DR2) of \Gaia~\cite{2018arXiv180409365G}.  
The network was trained on a combination of simulations and data.  The first round of training used the \mi galaxy from the \emph{Ananke} mock catalogs~\cite{2018arXiv180610564S}, based on the \Latte suite~\cite{Wetzel2016} of Milky Way-like galaxies from the \FIRE simulations~\cite{Hopkins:2013oba,2017arXiv170206148H}.  We used the star's formation history as defined by \cite{Necib:2018igl} to label a star as accreted if it was previously associated with a halo other than the host, and as \emph{in situ} otherwise. Transfer learning was then performed with Milky Way stars, using the \RAVE~DR5--\Gaia~DR2 cross match~\cite{2017AJ....153...75K} to label the most metal-poor stars ($\FeH < -1.5$) at vertical distances $|z| > 1.5\kpc$ as accreted for the training. 

The network takes as inputs the 5D kinematics of each star (two angular coordinates, two proper motions, and parallax) and then outputs a score associated with the probability that the star is accreted.  
When the network returns a score of $S=0$~($S=1$), this implies that the algorithm has maximal confidence that the star is \emph{in situ}~(accreted).  
We assign all stars with $S>0.75$ to the catalog; this cut is intended to maximize the purity of the sample while minimally biasing the resulting kinematic distributions, and is justified by extensive testing performed on mock catalogs of simulated galaxies~\cite{ml_paper}.  The network is optimized for and applied to
the subset of \Gaia~DR2 stars with parallax $\varpi > 0$ and fractional error $\delta \varpi/ \varpi < 0.10$. 

In this work, we focus on the subset of \Gaia~DR2 stars with radial velocity measurements that fall in the region of interest (ROI) defined by spherical Galactocentric radii $r \in [6.5, 9.5]\kpc$ and vertical distances $|z| < 3\kpc$.  We perform a Gaussian mixture analysis on the 3D velocities of all stars with a neural network score $S>0.95$. This sample consists of the stars that the network is most confident are accreted; focusing on this subset reduces the potential for disk contamination, although at the expense of possibly biasing the kinematic distributions \cite{ml_paper}.  A complete description of the mixture analysis, as well as other structures identified in the catalog, is provided in~ Necib et al. 2019, hereafter Paper I \cite{catalog_paper}. Here, we present evidence for a new stream discovered in this ROI, which comprises nearly 9\% of the stars in the sample with $S>0.95$.  We call this stream ``Nyx,'' after the Greek goddess of the night.  We have verified that Nyx does not match any previously catalogued stream in \cite{2010LNEA....4...13B,2016ASSL..420.....N, 2018MNRAS.474.4112M, 2018MNRAS.475.1537M, 2018ApJ...860L..11K}.

The left panel of Fig. 1 shows the Nyx stars in the Galactocentric spherical $v_r - v_\phi$ plane.  The dashed gray lines denote the 1, 2, and 3$\sigma$ contours enclosing the stars with scores $S<0.75$.  This subset is dominated by the \emph{in situ} stars, but also has contributions from accreted stars that are more difficult for the network to classify.  From Fig. 1, we see that the distribution is concentrated at $v_\phi \sim 220 \kms$ and $v_r \sim 0 \kms$.  The 2 and 3$\sigma$ contours drop down to $v_\phi \sim 0 \kms$ and non-zero $v_r$.  This effect is likely due to stars that belong to accreted structures with highly radial velocities (see~Paper I) that the network does not score as highly.
Nyx stars lie at the tail of the thick-disk distribution, but extend much farther out into the kinematic region associated with the stellar halo. 

\begin{figure}
\centering
\includegraphics[width=0.45\textwidth]{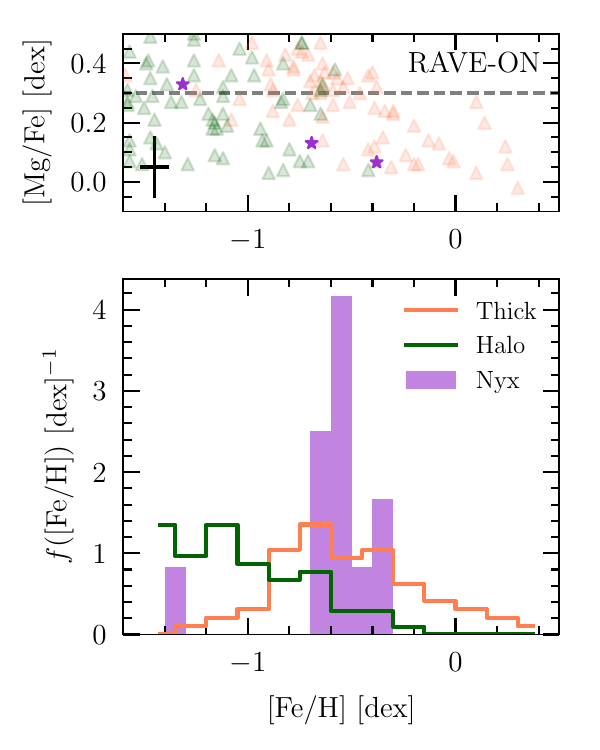}
\caption{\textbf{Chemical abundances of the Nyx stars.}  (Top) RAVE-ON abundances for the 3 Nyx stars with both [Mg/Fe] and [Fe/H] measurements.  The black cross shows the average measurement uncertainty.  (Bottom) Histogram of the 12 Nyx stars with RAVE-ON metallicities.  The chemical abundances of this small subset of Nyx stars suggest that it is relatively metal-rich and has [Mg/Fe]~$\lesssim 0.3$ (indicated by the gray dashed line). Overlaid are the distributions of the Milky Way halo and thick-disk stars from \cite{2004AJ....128.1177V};  we caution that these stars are classified using a  kinematic cut that would attribute Nyx to the thick disk.  The Nyx distributions are consistent with both a dwarf galaxy or thick-disk origin.  The tight clustering in chemical abundance space, if validated with further spectroscopic observations, would suggest a single progenitor origin.  }
\label{fig:alpha_fe}
\end{figure}

\begin{figure}
\centering
\includegraphics[width=0.45\textwidth]{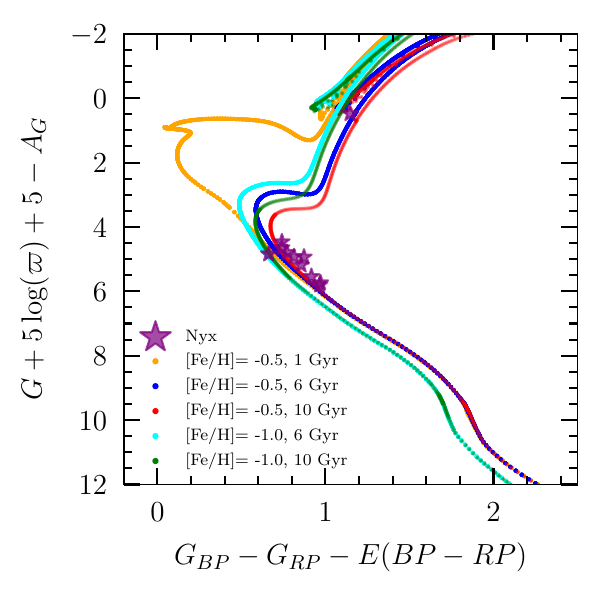}
\caption{\textbf{Color--Magnitude Diagram of the Nyx stars.}  As a reference, we show isochrones from \cite{2016ApJS..222....8D,2016ApJ...823..102C}, with $\FeH =-0.5$, $\FeH = -1$ and ages 1, 6, and 10 Gyr. The $x$-axis corresponds to the \Gaia colors $G_{BP} - G_{RP}$, dereddened by $E(BP-RP)$. The $y$-axis is the absolute magnitude in the $G$ band, corrected by the extinction $A_G$ \cite{2018A&A...616A...8A}.  We only show the 11 Nyx stars that pass the photometric cuts: \texttt{phot\_bp\_rp\_excess\_factor} $< 1.3 + 0.06~(G_{BP} - G_{RP})^2$ and $E(B - V) < 0.015$~\cite{2018A&A...616A..10G}. }
\label{fig:nyx_hr}
\end{figure}

The Nyx velocities are shown as arrows on a spatial projection of their locations in Fig. 1 (middle and right panels). In the following, we quote results in Galactocentric spherical coordinates.  This choice is made to be consistent with the modeling of \Gaia Enceladus in \cite{2019ApJ...874....3N}. Since this analysis is performed close to the disk plane, cylindrical and spherical coordinates are similar ($v_R \sim v_r$ and $v_z \sim v_\theta$).  For Nyx, the means of the best-fit multivariate normal velocity distribution are $\{\overline{v}_r, \overline{v}_\phi, \overline{v}_\theta \} = \left\{ 134.0^{+3.1}_{-3.7} , 130.0^{+2.4}_{-2.3}, 53.0^{+3.5}_{-3.6} \right\}\kms$, with dispersions $\{ \sigma_r, \sigma_\phi, \sigma_\theta \} = \left\{ 67.2^{+2.4}_{-2.5}, 45.8^{+1.8}_{-1.7}, 66.3^{+3.3}_{-3.0} \right\}\kms$.  Nyx has a significant non-zero radial velocity that causes its stars to move at an angle in the $x-y$ plane.  It is clearly prograde, moving with the Galactic disk, but lagging in velocity by $\sim$ 90 $\kms$. The 94 most likely stars to belong to  Nyx are coherent in velocity, with total average speed 282$\kms$ and dispersion of 32$\kms$. For reference, we provide a Toomre diagram of the Nyx stars in \SFig~1.  They are predominantly clustered within $\pm 1\kpc$ of the midplane, passing through the Solar neighborhood, but they also extend up to $\pm 2\kpc$.  Nyx stars can be found through roughly the full radial range studied. From Fig. 1, it appears that we are missing stars at $x \sim -9$, $y<0 \kpc$ and $x \sim -7$, $y > 0 \kpc$.  
We find that Nyx extends further in this direction when the neural network score cut is lowered to $S>0.85$ (see \SFigs~~ 2 and 3). Lowering the cut to $S>0.85$ does not entirely fill in this region, however.  It is likely that we are not capturing the entirety of the Nyx stream for a few reasons: the \Gaia selection function, the network mis-labeling Nyx stars with low scores, and the use of a Gaussian clustering algorithm to identify stars belonging to (what is likely an) inherently non-Gaussian structure.   Additionally, stars in this region have velocities pointing predominantly along the  line-of-sight, a direction our machine learning algorithm is insensitive to having only been trained on parallax, positions, and proper motions \cite{ml_paper}.
%

Few Nyx stars have accompanying spectroscopic measurements (see \SFig~4). 
Cross-matching with RAVE-ON~\cite{2017ApJ...840...59C} and applying a cut that the signal-to-noise ratio SNR $\geq$ 20, we find 12 stars that have measured [Fe/H] abundances, 3 of which also have [Mg/Fe] measurements.  As shown in Fig. 2,
the average value of [Fe/H] ([Mg/Fe]) is $-0.55^{+0.03}_{-0.04}$~($0.18^{+0.10}_{-0.14}$), with respective dispersion of $0.13^{+0.03}_{-0.04}$~($0.06^{+0.09}_{-0.05}$).  The errors quoted here are obtained by generating 100 iterations of abundances of each star, sampling over the errors in each measurement, then calculating the means and dispersions.  (The same figure using RAVE and GALAH data instead of RAVE-ON is shown as \SFig~5.)  For context, the average RAVE-ON measurement uncertainty on [Fe/H] ([Mg/Fe]) is 0.08~(0.11).  For the small subset for which we have data, Nyx stars have abundances that are comparable with both the thick disk (see Fig. 2) as well as with dwarf galaxies \cite{2004AJ....128.1177V,2009ARA&A..47..371T,2014MNRAS.444..515R}.  However, the small dispersion in the chemical abundances for Nyx does suggest a single progenitor origin, especially given the coherence in velocity space.  Given the small subset of Nyx stars with abundances, and the large measurement uncertainties from RAVE-ON, further spectroscopic follow-ups (for example from APOGEE-2, 4MOST, and WEAVE) are needed to validate these conclusions. 

\begin{figure*}
\centering
\includegraphics[width=0.90\textwidth]{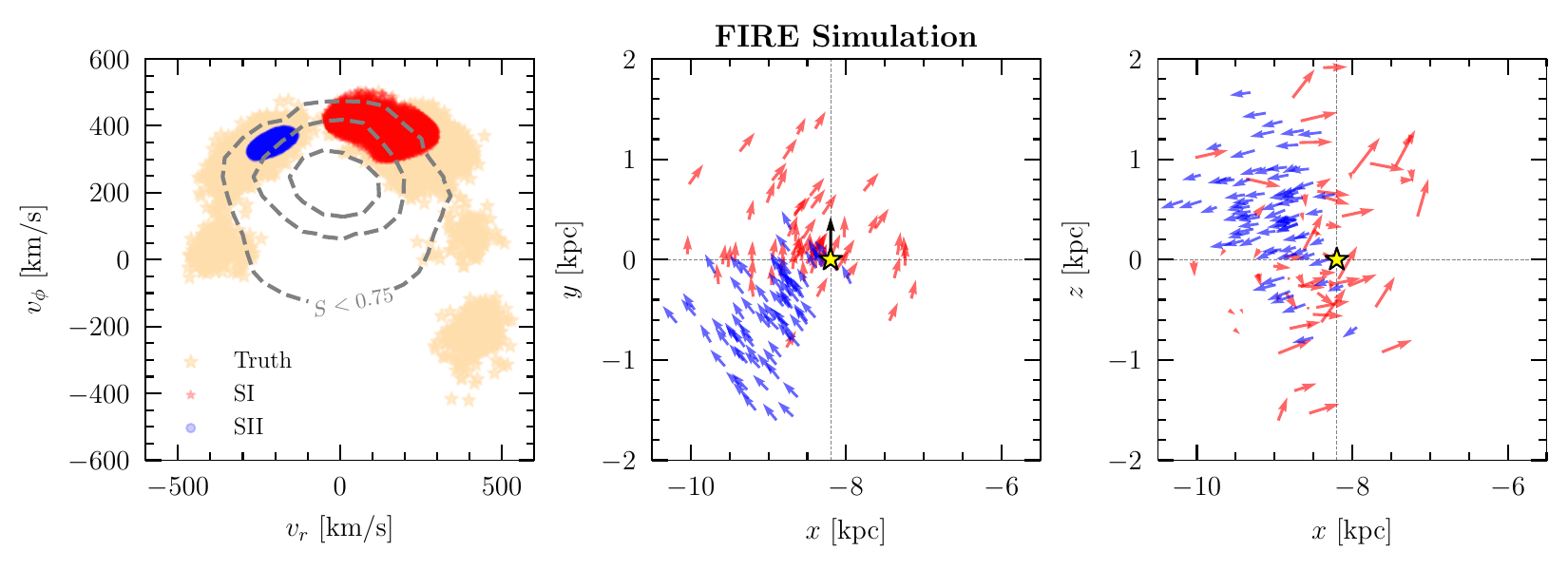}
\caption{\textbf{Example prograde stellar stream in a simulated Milky Way galaxy.  This figure is similar to Fig. 1,} but for a simulated Milky Way--mass galaxy from \FIRE and focusing on stars within $r_\odot < 2.3$ kpc of the sun.  In this case, the neural network training proceeds as before, except using the \mf galaxy as opposed to the actual Milky Way data. (Left)~Velocities of stars with $S>0.95$ that belong to the \texttt{m12f} merger which leaves the most significant amount of stellar debris at the LSR (orange points). The red and blue points show the two subsets of stars that are identified by the Extreme Deconvolution clustering  algorithm --- we refer to these as streams SI and SII, respectively. See the main text for a discussion justifying why only a part of the stream is selected using this procedure.  (Middle/Right) Velocity distributions in the $x - y$ and $x - z$ planes for the stars that are associated with SI and SII.  These streams are concentrated in the Galactic plane and are prograde, similar to Nyx. 
  }
\label{fig:m12f}
\end{figure*}

In Fig. 3, we show the Nyx stars in the Hertzsprung--Russell (HR) diagram, in the \Gaia bands $G_{BP} - G_{RP}$ and absolute magnitude $G + 5 \log(\varpi) + 5$. We corrected for the reddening $E(BP-RP)$ and extinction $A_G$ using the associated \Gaia values from~\cite{2018A&A...616A...8A}.  Due to possible contamination from the backgrounds, we choose stars with \texttt{phot\_bp\_rp\_excess\_factor} $< 1.3 + 0.06~(G_{BP} - G_{RP})^2$, and only select low-extinction stars with $E(B - V) < 0.015$ \cite{2018A&A...616A..10G}. Only 11 Nyx stars pass these photometric cuts.  For reference, we overlay three isochrones with metallicities $\FeH = -0.5$, -1,  and stellar ages 1, 6, and 10~Gyr from the MIST project \cite{2016ApJS..222....8D,2016ApJ...823..102C}.  The few Nyx stars that pass these photometric cuts are consistent with the older isochrones. This strengthens the case that Nyx originated from a disrupted satellite galaxy, because disk stars span a larger range of metallicities and ages and tend to be younger (see \eg \cite{martig2016radial}).
  
We now turn to estimating the orbital parameters of Nyx stars. To do so, we use the \texttt{gala} package \cite{gala} and assume the Milky Way potential of~\cite{2015ApJS..216...29B}.  Running the orbits of the stars backwards in time by 1~Gyr, we find that the mean eccentricity is $0.60$ (dispersion of $0.12$), mean pericenter is $3.9\kpc$, mean apocenter is $16\kpc$, and mean $z_\text{max}$ is $1.7\kpc$ (see Supplementary Figs.~6 and~7).  These orbital properties are distinct from what is expected for the thin disk, which has eccentricities close to circular.  The orbits are also more eccentric than expected of the thick disk, which has eccentricities peaked at $e\sim 0.3$ and concentrated below $e\sim 0.5$~\cite{2017ApJ...850...25L}. In \SFig~7, we show the orbits traced by the Nyx stars integrated back by 1 Gyr; the prograde and highly eccentric motion of the Nyx stars is evident. Coupling this observation with the fact that Nyx lags behind the disk by $\sim 90\kms$ and has a significant radial velocity component makes a strong case that it is the result of a satellite merger.  

Stellar ``moving groups'' can arise from resonances in stellar orbits that are perturbed by the Galactic bar~\cite{Dehnen_2000, Fux:2001sm}  
or density waves caused by the spiral arms~\cite{2004MNRAS.350..627D, Quillen:2005uk}.  However, simulations of bar and spiral arm perturbations typically result in coherent velocity structures that lag the Galactic disk by only $\sim 20$--$50\kms$~\cite{2010LNEA....4...13B}.
It is in principle conceivable that a stronger perturbation to the disk could have been indirectly induced by the passage of a satellite galaxy through the disk, without necessarily leaving many stars behind: the dynamical ``kick'' caused by such a passage can result in density waves moving in the same direction as the Sun at velocities comparable to what we find for Nyx~\cite{2009MNRAS.396L..56M, 2009MNRAS.397.1599Q}.  However, such perturbations generically result in a specific shape (``arches'') in the \textit{cylindrical} $v_R - v_\phi$ plane. 
This effect has been observed in \Gaia~DR2~\cite{2018A&A...616A..11G}, and does not resemble the kinematics of Nyx.  Indeed, as will be discussed later, there are hints for additional members of the Nyx stream that may fall along constant eccentricity surfaces~\cite{Helmi:2005wc}, as expected of a satellite merger, rather than constant energy surfaces, as expected for a disk perturbation~\cite{2009MNRAS.397.1599Q}.  

Taking this evidence together, we argue that Nyx is a remnant of a dwarf galaxy.  Its characteristics are largely consistent with previous studies of accreted stellar disks~\cite{2008MNRAS.389.1041R, 2009MNRAS.397...44R, 2009ApJ...703.2275P, 2017MNRAS.472.3722G}.  Namely, it is corotating with a lag speed that falls in the range motivated by  simulations.  Additionally, its mean metallicity of $\FeH~\sim~-0.55$ is consistent with a galaxy of stellar mass in the range $[4 \times 10^9, 8 \times 10^{9}] \Msun$, as estimated using the mass-metallicity relation of~\cite{2013ApJ...779..102K}.  Note that we account for both the theory and RAVE-ON measurement uncertainties in this estimate. The range quoted is bracketed by the 16 and 84 percentiles.  The lower end of this range is compatible with the mass scale of the satellites studied in~\cite{2008MNRAS.389.1041R, 2009MNRAS.397...44R, 2009ApJ...703.2275P,2017MNRAS.472.3722G} that led to viable accreted stellar and dark disks. In general, disk dragging of satellites is most efficient for such massive systems.  However, we do caution that this is a simple estimate, and does not account for the potential metallicity gradients that would be expected as various layers of stars are tidally stripped from the satellite progenitor. Dedicated simulations of different merging scenarios will establish the feasibility of Nyx, either as an accreted satellite or a thick disk perturbation.

To demonstrate that such metal-rich prograde streams can exist in galaxies like our own --- and that our analysis procedure correctly identifies them if they do --- we repeat our study on the mock catalog of the \FIRE simulation \mf, focusing on the region within $2.3\kpc$ of LSR1~\cite{2018arXiv180610564S}. The general properties and merger history of the \mf galaxy are described in~\cite{2018ApJ...869...12S,Necib:2018igl}. Proceeding in a parallel manner to our \Gaia study, we train the network on the \mi galaxy, then apply transfer learning on \mf stars. It is important to note that \mi and \mf are completely different (but both Milky Way-mass) galaxies from different regions in a cosmological volume. Their merger histories, in particular, are quite distinctive: the evolution of \mf is more active up to redshift $\sim 1$, while \mi is more quiescent at late redshifts \cite{2017arXiv170206148H}.

To identify individual populations in velocity space, we run the Extreme Deconvolution mixture analysis on the subset of \mf stars with network scores $S>0.95$ using four Gaussian distributions. Note that this is different from the method used to identify Nyx (see Paper I).  It is computationally faster, but does not yield uncertainties on the best-fit parameters, which we do not need for this case. 
Two Gaussians (shown in blue and red on the left panel of Fig. 4) are matched with the largest stream of \mf, which was labeled as Merger~I in \cite{Necib:2018igl}. 
We refer to these as SI and SII, respectively.  The progenitor is a massive satellite that merged at redshift $\sim 0.12$--0.39, with a peak dark matter halo mass of $1.5\times10^{11}\Msun$, and a total stellar mass right before merger of $1.8 \times 10^9\Msun$.  Its tidal debris has an average stellar metallicity of $\FeH \sim -0.88$ \cite{Necib:2018igl}, alpha abundance $\alphaFe = 0.26$, and stellar age of $9.6$~Gyr.  

The right two panels of Fig. 4 show the spatial distribution of the velocities for SI and SII as  red and blue arrows, respectively.  Similarly to Nyx, both streams corotate with the disk and are concentrated within a few kpc of the midplane.  Both SI and SII are highly radial, but move in opposite directions towards/away from the galactic center.  This suggests that they correspond to two passages of the stream near the solar position in \mf.  The total velocity of SI (SII) is 440 (419) km/s with a dispersion of 29 (11) km/s. Future studies of the simulations will unveil the different possible stream velocities and dispersions expected from such mergers, which will enable us to better understand the origin of Nyx.

We can also check if there are other stars in the $S>0.95$ sample that originated from the same progenitor as SI and SII, but that are not identified by the mixture analysis.  These stars are shown in pale orange in the left panel of Fig. 4.  Notably, SI and SII comprise only a fraction of the stream in the $S>0.95$ sample, likely because the mixture analysis separates the individual populations assuming Gaussian velocity distributions, and the stream distribution is inherently non-Gaussian.  SI and SII comprise an even smaller fraction of the \emph{total} stream, because demanding a score cut of $S>0.95$ significantly reduces the number of stream stars in the sample; only $19\%$ of the \mf stream stars pass this cut.  This fraction increases to 41\% when reducing the cut on the network score to $S>0.85$, though at the expense of introducing more disk contamination to the sample. Looking at the purity of the system (defined as the fraction of stars passing the score cut that are truly accreted), we find that it is $53\%$ for $S>0.95$ but drops to 43\% for $S > 0.85$.

The example from the \mf simulated galaxy serves as a proof-of-principle that not only a prograde stream can comprise a non-trivial fraction of the accreted stars near the Solar neighborhood, but also that accreted stars from the disk can be identified using this method.  This test case also suggests that additional remnants of the Nyx stream may exist that are not captured in the $S>0.95$ sample, or by the mixture model analysis.  To test the first hypothesis, we repeat our study using the $S>0.85$ sample of accreted stars (see~Paper I for details). 
We find a population of stars with properties consistent with Nyx, as well as an additional stream (referred to as Nyx-2) with similar prograde motion and chemical abundances, but opposite radial velocity.  This suggests that Nyx-2 is related to Nyx and might actually be debris from a separate passage of the same satellite, similar to the \mf example above (see \SFig~2). Dedicated simulations will be important in unveiling the origin of Nyx, to identify the mass and infall time of the progenitor if it is a merging object, and its subsequent effect on the disk.

Follow-up spectroscopic studies will play a crucial role in establishing the origin of the Nyx stream by providing a larger set of chemical abundances. This will help either confirm the kinematic arguments above that suggest that Nyx is the remnant of a dwarf galaxy, or that it is a perturbation of the Milky Way disk. Additionally, if Nyx-2 and Nyx have similar elemental abundances, it would substantiate their relation to each other.  Our study of a simulated Milky Way-like galaxy demonstrates that infalling dwarf galaxies can leave signatures very similar to what we observe for Nyx. 
If Nyx is indeed the result of such a merger, then it would provide evidence for accreted prograde stars, and potentially, an accompanying dark matter component~\cite{Necib:2018igl} in a stream or disk.  The presence of such a dark matter component would substantially alter our current understanding  of the local dark matter phase-space distribution, and have important ramifications for terrestrial searches for the dark matter particle.

\section*{Data Availability}

The accreted star catalog used for this analysis is available at \url{https://doi.org/10.5281/zenodo.3579379}. The simulation \mf is available at \url{https://fire.northwestern.edu/ananke/}. The IDs of Nyx stars are available as part of the Supplemental Material of this work. 

\section*{Code Availability}

This analysis makes use of \emph{emcee} and Extreme Deconvolution for the Gaussian Mixture model. The Python Markov chain Monte Carlo code \texttt{emcee} is freely available and documented at \url{http://dfm.io/emcee/current/}. Extreme Deconvolution is freely available at \url{https://github.com/jobovy/extreme-deconvolution}. Details regarding the application of these two public codes are provided in \cite{catalog_paper}.

\section*{Author Contribution}

All authors discussed the results and commented on the manuscript.  ML and TC conceived the project.  LN built the data analysis pipeline.  BO, TC, and MF conceptualized the machine learning algorithms.  BO built the deep neural network and produced the accreted stellar catalog. Interpretation of the results and writing of the original manuscript were done by LN and ML.  The FIRE-2 simulation code was built by PFH, and run by PFH, SGK, and AW. SGK and AW ran the halo finding algorithm on the simulation, and LN identified the mergers as a function of redshift. RS built the mock catalogs used in the training of the neural network.

\section*{Corresponding Author}

Correspondence to Lina Necib \textbf{(lnecib@caltech.edu)}.

\section*{Acknowledgments}

We thank A.~Helmi, J.~Johnson, E.~Kirby, N.~Laracy, J.~Read, N.~Shipp, and J.~Wojno for helpful discussions.  This work was performed in part at Aspen Center for Physics, which is supported by National Science Foundation grant PHY-1607611.  This research was supported by the Munich Institute for Astro- and Particle Physics (MIAPP) of the DFG cluster of excellence ``Origin and Structure of the Universe."  This research was supported in part by the National Science Foundation under Grant No. NSF PHY-1748958.

LN is supported by the DOE under Award Number
DESC0011632, and the Sherman Fairchild fellowship.
ML is supported by the DOE under contract DESC0007968 and the
Cottrell Scholar Program through the Research Corporation for Science Advancement. 
BO and TC are supported by the US Department of Energy under grant number DESC0011640.  
MF is supported by the Zuckerman STEM Leadership Program and in part by the DOE under grant number DE-SC0011640.
SGK and PFH are supported by an Alfred P. Sloan Research Fellowship, NSF Collaborative Research Grant \#1715847 and CAREER grant \#1455342, and NASA grants NNX15AT06G, JPL 1589742, 17-ATP17-0214.
AW is supported by NASA, through ATP grant 80NSSC18K1097 and HST grants GO-14734 and AR-15057 from STScI, and a Hellman Fellowship from UC Davis.
This work utilized the University of Oregon Talapas high performance computing cluster.
Numerical simulations were run on the Caltech compute cluster ``Wheeler,'' allocations from XSEDE TG-AST130039 and PRAC NSF.1713353 supported by the NSF, and NASA HEC SMD-16-7592.

RES thanks Nick Carriero, Ian Fisk, and Dylan Simon of the Scientific Computing Core at the Flatiron Institute for their support of the infrastructure housing the synthetic surveys and simulations used for this work.

This work has made use of data from the European Space Agency (ESA) mission Gaia (http://www.cosmos.esa.int/gaia), processed by the Gaia Data Processing and Analysis Consortium (DPAC, http://www.cosmos.esa.int/web/gaia/dpac/consortium). Funding for the DPAC has been provided by national institutions, in particular the institutions participating in the Gaia Multilateral Agreement.

Funding for RAVE has been provided by: the Australian Astronomical Observatory; the Leibniz-Institut fuer Astrophysik Potsdam (AIP); the Australian National University; the Australian Research Council; the French National Research Agency; the German Research Foundation (SPP 1177 and SFB 881); the European Research Council (ERC-StG 240271 Galactica); the Istituto Nazionale di Astrofisica at Padova; The Johns Hopkins University; the National Science Foundation of the USA (AST-0908326); the W. M. Keck foundation; the Macquarie University; the Netherlands Research School for Astronomy; the Natural Sciences and Engineering Research Council of Canada; the Slovenian Research Agency; the Swiss National Science Foundation; the Science $\&$ Technology Facilities Council of the UK; Opticon; Strasbourg Observatory; and the Universities of Groningen, Heidelberg and Sydney.
The RAVE web site is at \url{https://www.rave-survey.org}.

\section{Methods}
A detailed presentation of the analysis methods is provided in two companion papers.  For a discussion of the machine learning methods, including critical validation tests, see~\cite{ml_paper}.  For a discussion of the clustering methods and further interpretation of the results when applied to the accreted star catalog, see~\cite{catalog_paper}. 

\def\bibsection{}
\bibliographystyle{naturemag}
\bibliography{nyx}


\appendix

\onecolumngrid
\pagebreak
\clearpage
\section{Supplemental Material}

\setcounter{equation}{0}
\setcounter{figure}{0}
\setcounter{table}{0}
\setcounter{section}{0}
\makeatletter
\renewcommand{\theequation}{S\arabic{equation}}
\renewcommand{\thefigure}{S\arabic{figure}}
\renewcommand{\thetable}{S\arabic{table}}

This analysis makes use of \emph{emcee} \cite{2013PASP..125..306F} and Extreme Deconvolution \cite{2011AnApS...5.1657B} for the Gaussian Mixture model.

 \begin{figure*}[h]
 \centering
 \includegraphics[width=0.45\textwidth]{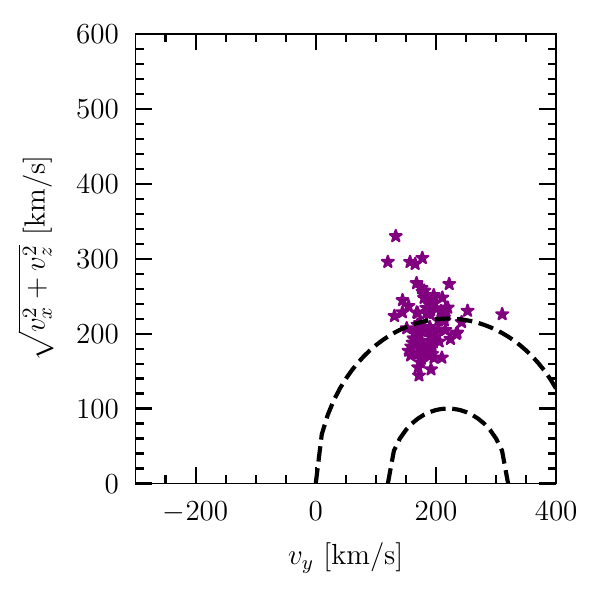}
 \caption{\textbf{Toomre diagram for Nyx stars.}  For reference, the black dashed lines correspond to velocity contours of 100 and 220~km/s in the local frame. }
\label{fig:nyx_toomre}
 \end{figure*}

\begin{figure*}[h]
\centering
\includegraphics[width=0.90\textwidth]{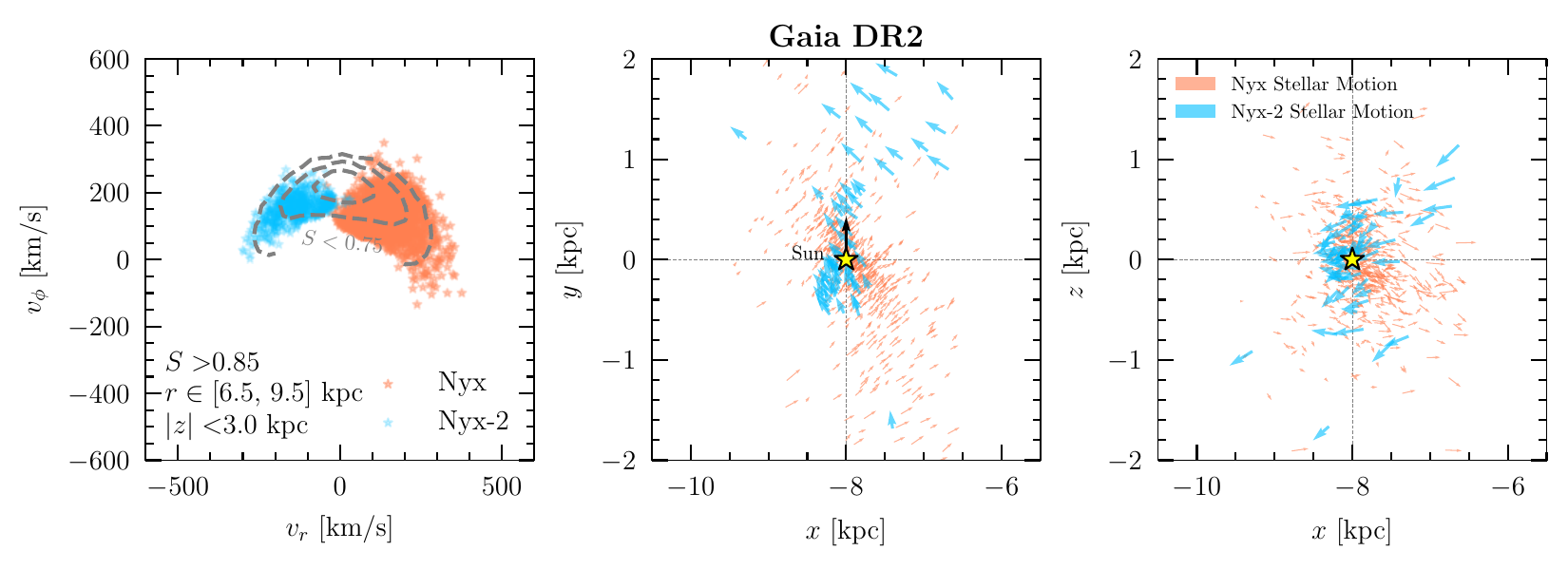}
\caption{\textbf{Dynamics of the Nyx and Nyx-2 stars.  This figure is similar to Fig. 1,} but for stars satisfying a network score cut of $S>0.85$. In this case, we identify an additional group of stars, labeled Nyx-2, that may be related to the Nyx stream.  Details of the clustering algorithm used to find Nyx and Nyx-2 are provided in~Paper I. (Left) Plot of the velocities of Nyx and Nyx-2 stars, in the Galactocentric spherical $v_r - v_\phi$ plane.  (Middle/Right) Velocity of Nyx and Nyx-2 stars in the $x - y$ and $x - z$ planes, where $x - y$ is the disk plane, and $z$ is perpendicular to the disk.  Like Nyx, Nyx-2 is prograde and radial.  Its stars move away from the Galactic Center, rather than towards, suggesting that it may be debris from a separate passage of the same satellite galaxy.   }
\label{fig:toomre_075}
\end{figure*}

\vspace{0.5in}

\begin{figure*}[h]
\centering
\includegraphics[width=0.90\textwidth]{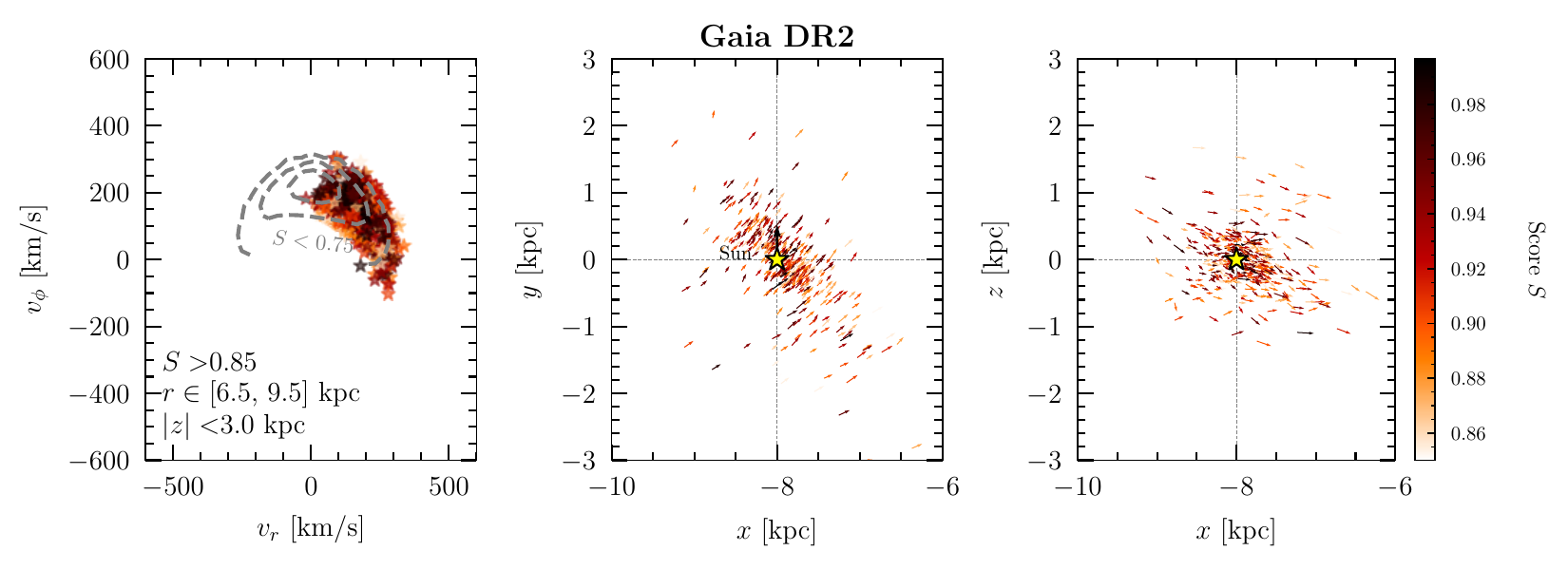}
\caption{\textbf{Network score dependence of the Nyx stars.  This figure is similar to Fig. 1,} but for the Nyx stars in the $S>0.85$ sample.  Stars are colored by their neural network score.}
\label{fig:toomre_075_colored}
\end{figure*}

\vspace{0.5in}

\begin{figure*}[h]
\centering
\includegraphics[width=0.7\textwidth]{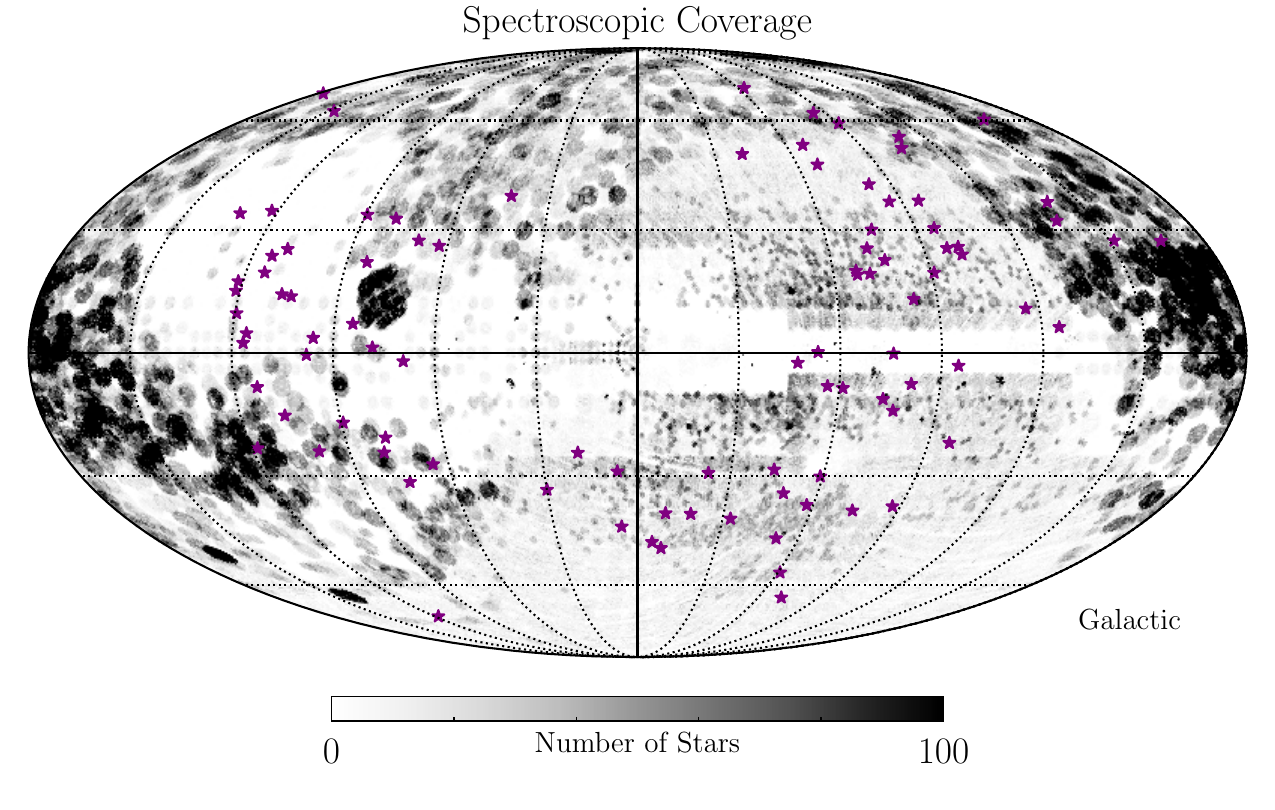}
\caption{\textbf{Skymap of the Nyx stars.  This figure compares}  Nyx (purple stars) to the footprint of current spectroscopic surveys (gray) using data from \cite{2018MNRAS.481.4093S}. The colorbar indicates the number of stars, which we saturate at 100 for clarity. Nyx has the greatest overlap with the RAVE-ON catalog~\cite{2017ApJ...840...59C}. }
\label{fig:spec_coverage}
\end{figure*}

\begin{figure}[h]
\includegraphics[width=0.45\textwidth]{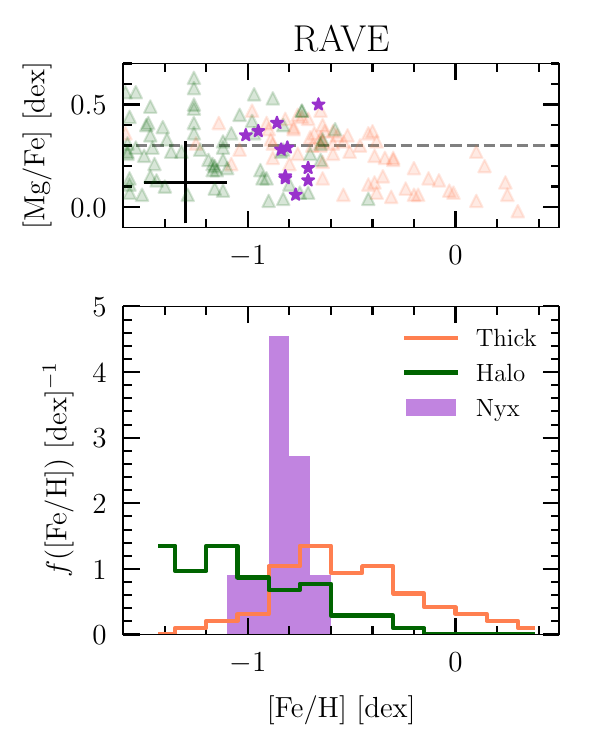}
\includegraphics[width=0.45\textwidth]{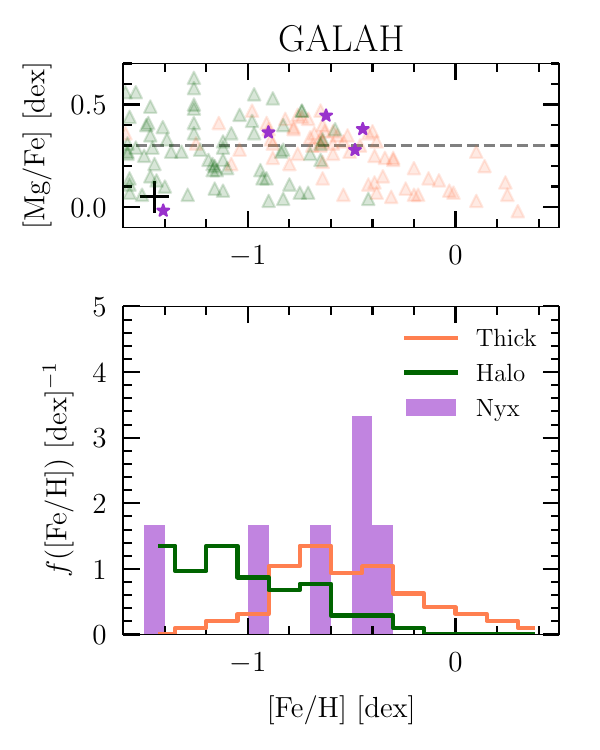}
\caption{\textbf{Chemical abundance of the Nyx stars.  This figure is similar to Fig. 2},  but with abundances from RAVE~\citep{2017AJ....153...75K}~(left) and GALAH~\citep{2018MNRAS.478.4513B}~(right), rather than RAVE-ON.  The error bars are the 1-sigma uncertainties of the surveys. }
\label{fig:alpha_fe_rave}
\end{figure}

 \begin{figure*}[h]
 \centering
 \includegraphics[width=0.9\textwidth]{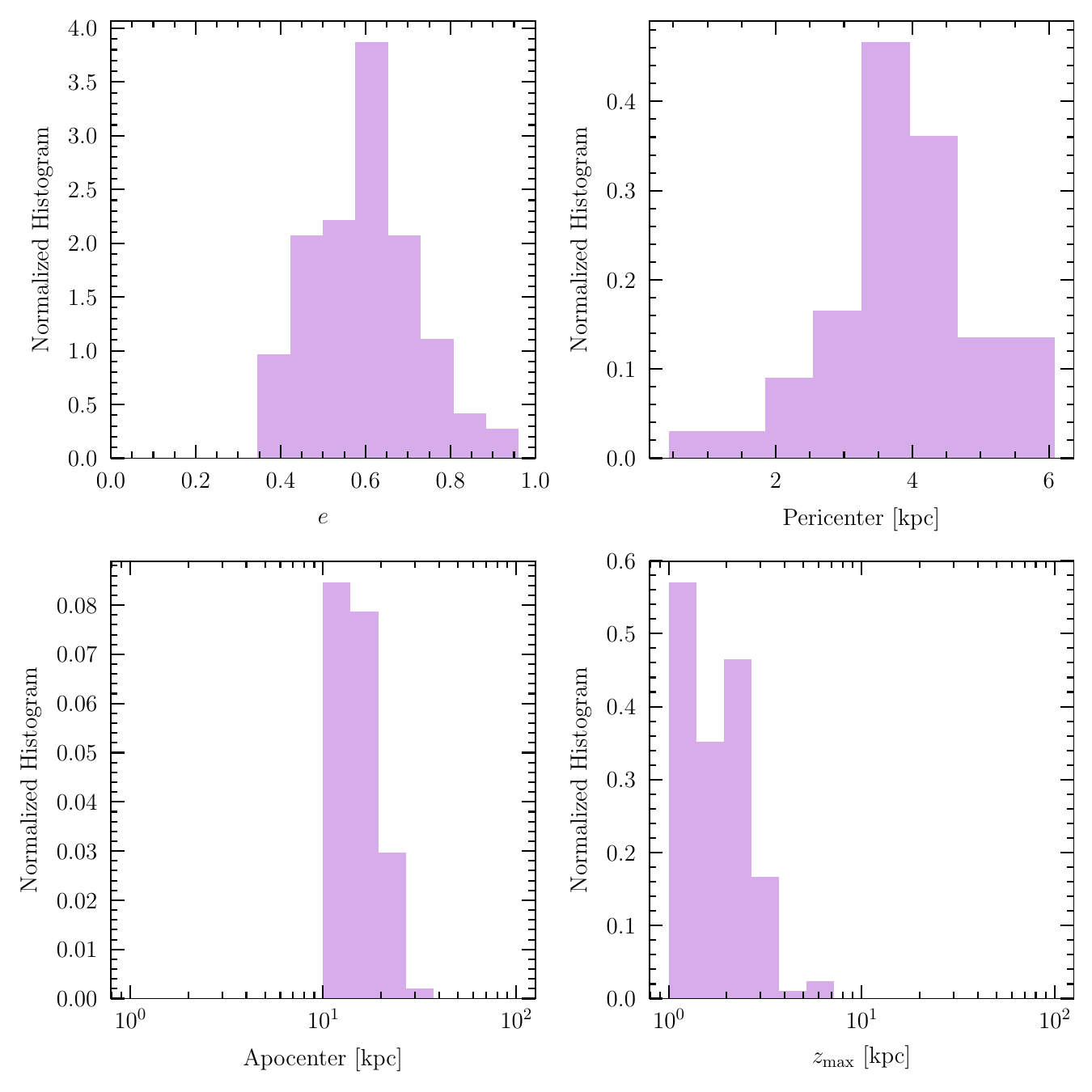}
 \caption{ \textbf{Orbital parameters of the Nyx stars.}  The distribution of eccentricities (top left), pericenters (top right),  apocenters (bottom left), and maximum vertical distance, $z_\text{max}$, from the Galactic plane (bottom right) for the Nyx stars.  These parameters were obtained running the orbits of Nyx stars backwards in time over a period of 1 Gyr using the \texttt{gala} package \citep{gala} and the Milky Way potential of  \cite{2015ApJS..216...29B}.  }
 \label{fig:nyx_orbits}
 \end{figure*}

 \begin{figure*}[h]
 \centering
 \includegraphics[width=0.55\textwidth]{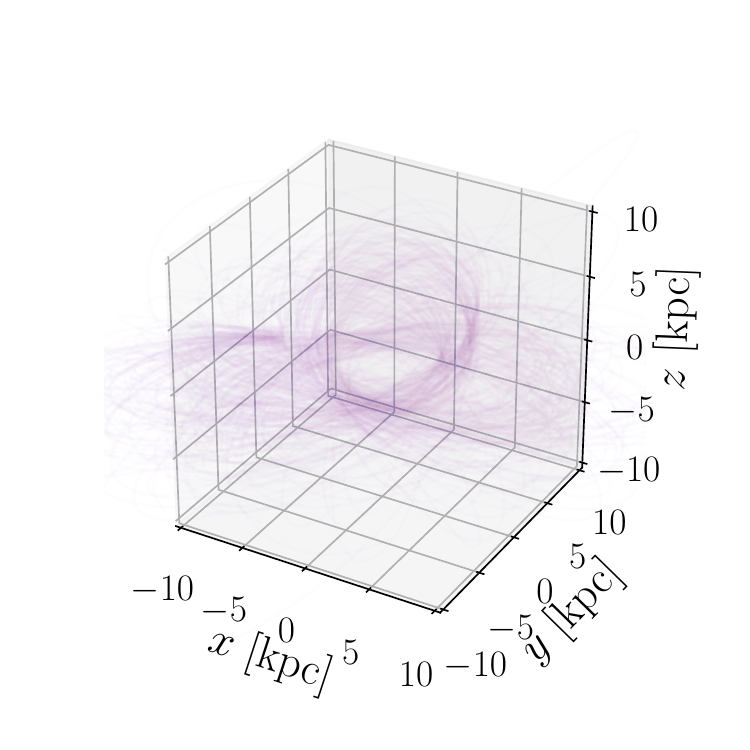}
 \caption{ \textbf{The orbits of the Nyx stars.  These are obtained by integrating} back by 1 Gyr using the \texttt{gala} package \citep{gala} and assuming the Milky Way potential of  \cite{2015ApJS..216...29B}. }
\label{fig:nyx_backwards}
 \end{figure*}

\end{document}